\documentclass[english]{iopart}
\usepackage{amsbsy}
\usepackage{babel}
\pdfoutput=1
\usepackage{graphicx,color}

\usepackage{latexsym,amssymb}
\usepackage{graphics}

\begin{document}

\title{Models with short and long-range interactions: phase diagram and reentrant phase}

\author{Thierry Dauxois $^1$, Pierre de Buyl $^2$, Leonardo Lori $^{1,3}$ and Stefano Ruffo $^{1,3}$}

\address{1. Laboratoire de Physique de l'{\'E}cole Normale Sup{\'e}rieure de
Lyon, Universit{\'e} de Lyon, CNRS, 46 All{\'e}e d'Italie, 69364 Lyon
c{\'e}dex 07, France
\\
2. Center for Nonlinear Phenomena and Complex Systems, Universit{\'e} Libre de
Bruxelles (U.L.B.), Code Postal 231, Campus Plaine, B-1050 Brussels, Belgium \\
3. Dipartimento di Energetica ``S. Stecco", Universit{\`a} di Firenze and INFN, via
s. Marta, 3, 50139 Firenze, Italy}



\begin{abstract}
We study the phase diagram of two different Hamiltonians 
with competiting local, nearest-neighbour, and mean-field couplings. The first example corresponds to 
the HMF Hamiltonian with an additional short-range interaction.
The second example is a reduced Hamiltonian for dipolar layered spin structures, with a new feature with 
respect to the first example, the presence of anisotropies.
 
The two examples are solved in both the canonical and the microcanonical ensemble using a combination
of the min-max method with the transfer operator method. The phase diagrams present typical features 
of systems with long-range interactions: ensemble inequivalence, negative specific heat and temperature jumps. 

Moreover, in a given range of parameters, we report the signature of phase reentrance. This can also be interpreted as the presence 
of azeotropy with the creation of two first order phase transitions with ensemble inequivalence, as one parameter is
varied continuously.

\end{abstract}

\maketitle

\section{Introduction}
Interactions among elementary constituents of matter can be classified according to the features of the
two-body potential. If this latter decays at large distance with an exponent that is bigger than space
dimension, one speaks of short-range interactions: otherwise interactions are long-range 
\cite{review,longrange,assisi,leshouches}. The difference between these two types of interactions has consequences 
at both the thermodynamic
and dynamical level. In particular, systems with long-range interactions present a series of peculiar properties
which are mainly a consequence of the lack of additivity of energy. Perhaps, the most impressive one is
{\it ensemble inequivalence} \cite{BMR01}, which entails the presence of negative specific heat in the microcanonical
ensemble. Recently, there has been a growing interest in systems with long-range interactions. Besides realizing that
they appear in many branches of physics (self-gravitating systems, two-dimensional hydrodynamics, dipolar
interactions, unscreeened plasmas, etc.), the main breakthrough has been the discovery that simple
toy models of the mean-field type reproduce many of the fundamental properties of systems with long-range
interactions. These models can be solved exactly in both the canonical and the microcanonical ensemble, therefore
ensemble inequivalence can be checked explicitly. Moreover, the dynamics of Hamiltonian systems with
long-range interactions is well described by the
Vlasov equation \cite{Nicholson}, which allows one to interpret the presence of quasi-stationary states
as stable stationary states of the Vlasov equation.

An interesting but largely unexplored aspect of the physics of systems with long-range interactions is the combined effect of
terms of short and long-range type. The Ising chain with mean-field (Curie-Weiss) interaction and
nearest-neighbour interactions has been solved in the canonical ensemble by Nagle \cite{nagle} and Kardar
\cite{kardar} (see also Ref.~\cite{russe}). The complete microcanonical solution of this model has been recently obtained \cite{MRS05}
and it has been realized that the addition of a short-range term to the Curie-Weiss one is enough to
generate ensemble inequivalence. The emphasis of this latter paper was rather on {\it ergodicity breaking},
a property which is induced by the lack of {\it convexity} of the accessible region of macroscopic thermodynamic
parameters (here energy and magnetization). It was later realized that ergodicity breaking can be found 
generically in systems with long-range interactions \cite{BDMR08,review}. 

After the introduction of an XY model with mean-field interactions \cite{HMF}, the so-called HMF model, 
a model defined on a one-dimensional
lattice and with additional nearest-neighbour interactions has been considered \cite{Campa}. A preliminary
study of the phase diagram and of the robustness of quasi-stationary states has been performed. As for the
previous Ising-type model, ensemble inequivalence is induced by the addition of the short-range term, since inequivalence
is absent in the pure mean-field XY model. In addition to results already published elsewhere \cite{Campa,review}, we 
present here an elegant method to derive the second order phase transition line using  a perturbation theory inspired by quantum mechanics.

There is a growing interest in finding physical systems in which it could be possible to observe experimentally 
the effects found in systems with long-range interactions. Examples are the free electron laser and the 
collective atomic recoil laser discussed in Ref.~\cite{bachelard} in this issue, which should allow one to
observe quasi-stationary states. 

For a certain class of layered magnets (e.g. $(C H_{3}NH_3)_2CuCl_4$),  it has been remarked \cite{campionedip} that,
in specific temperature ranges and for given shapes of the samples, it is possible to reduce the
microscopic Hamiltonian of the system to that of an effective Hamiltonian of classical rotators on a
one-dimensional lattice with local, nearest-neighbour and mean-field interactions. Indeed, this Hamiltonian turns 
out to be very similar to that introduced in Ref.~\cite{Campa}. This reduction is generic
for systems dominated by dipolar forces \cite{Bramwell,landau}, for which the energy is marginally superextensive.
Indeed, it turns out that an extensive energy can always be obtained, but it contains a shape
dependent ``demagnetizing'' term proportional to the square of the magnetization. In our case, this
latter term has the form of a mean-field Curie-Weiss type interaction~\cite{Layer}.

In this paper, we will determine the phase diagram of this effective Hamiltonian for different values
of its parameters. We will also take the opportunity to explain in detail the method of solution in
both the canonical and microcanonical ensemble. It consists of an interesting combination of the
{\it min-max} method introduced in Ref.~\cite{Leyvraz} with the transfer integral 
method \cite{krum,Aubry,solitons}. As for the transfer integral, we will apply the Bode discretization method \cite{Bode},
which allows us to obtain a much more precise integration of the transfer operator with respect to
previous studies \cite{Campa}.

Moreover, a phenomenon of {\it phase reentrance} \cite{fasirientranti,staniscia} is observed 
for the model we discuss in this paper. When lowering the temperature, one observes a disorder/order
transition followed by an order/disorder one, in such a way that the low temperature phase has zero
magnetization. 

The paper is organized as follows. In Section~\ref{HMF}, we introduce and show explicitly the solution 
of the XY model with short and long-range interactions \cite{Campa}. In Section~\ref{reduced}, we study
the phase diagram of the reduced Hamiltonian which describes the layered magnetic system and discuss
the phenomenon of phase reentrance. In Section~\ref{conclusions}, we draw some conclusions.

\section{HMF model with additional short-range interactions}
\label{HMF}

Let us consider the Hamiltonian~\cite{review,Campa}
\begin{equation}
\label{eq:campa}
H=\sum_{i=1}^N \frac{p_i^2}{2}-K \sum_{i=1}^{N}\cos\left(\theta_{i+1}-\theta_i\right) 
+ \frac{J}{2N}\sum^N_{i,j=1}[1-\cos(\theta_i-\theta_j)]
\end{equation}
where $(\theta_i,p_i)$ are canonical conjugate variables, with $\theta_i$ representing an angle
on the unit circle. This Hamiltonian models a 1D lattice whose sites are occupied by rotors: the lattice
has periodic boundary conditions $\theta_{i+N}=\theta_i$. Two kinds of interaction are present: one
with coupling $K$, which is nearest-neighbour, and one with ferromagnetic coupling $J$, which is 
of mean-field type. We will allow the coupling $K$ to take both positive and negative values.
When $K=0$, one recovers the Hamiltonian Mean Field (HMF) model, whose phase diagram in both the
canonical and microcanonical ensemble has been widely studied. The model displays a second order
phase transition at the critical temperature $T_c=J/2$ and the corresponding critical energy
$u_c=3J/4$. Conversely, when $J=0$, the model reduces to the well known chain of coupled rotators,
which may have interesting dynamical effects \cite{EscandeRuffo} but does not have phase transitions.
When combined, the two interaction terms produce a very rich phase diagram with first and second order
phase transitions and a tricritical point. 
For convenience, since we will always use ferromagnetic mean-field interactions all along this
paper, such as to generate order in 1D, we will set $J=1$. It has been shown \cite{Campa} that ensembles are not
equivalent for this model and that the canonical phase diagram differs from the microcanonical
one. All the interesting features of models with long-range interactions are present: negative specific heat,
temperature jumps, breaking of ergodicity \cite{review}. Here, we will not review these
results, but we will introduce the method which allows to solve the model, because it will be used
also for the variant of model (\ref{eq:campa}) that we present in this paper. Our theoretical approach
combines three methods: one is specific of short-range interactions, the transfer integral method
\cite{krum,Aubry,solitons}; the second is standard for mean-field models, the saddle point method \cite{Angel}; 
the third, the min-max method \cite{Leyvraz}, has been more recently introduced and
allows one to get microcanonical entropy from the canonical free energy.
Other solution methods introduced for long-range interactions, like the large-deviation method
\cite{BarreJstat,Touchette}, cannot be simply extended to this class of model because of the short-range
component of the Hamiltonian, which does not allow one to write the Hamiltonian in terms of
``global'' variables \cite{review}.

The phase transitions we will discuss in this paper are all characterized by the XY magnetization   
\begin{equation}
\mathbf{m}=\frac{1}{N}\left(\sum_{i=1}^N\cos\theta_i,\sum_{i=1}^N\sin\theta_i\right)=\left(m_x,m_y\right),
\end{equation}
which is a vector whose modulus $m$ is finite in the broken symmetry phase.

In order to derive the equilibrium properties of model (\ref{eq:campa}), one first computes the partition
function in the canonical ensemble as follows
\begin{eqnarray}
Z &=&\int \prod_{i=1}^N (\mbox{d}\theta_i \mbox{d}p_i)\exp(-\beta H)\\
&=&\left(\frac{2\pi}{\beta} \right)^{\frac{N}{2}}\exp\left(-\frac{\beta N}{2}\right)
\int \prod_{i=1}^N \mbox{d}\theta_i \exp \left( \frac{\beta}{2N}\left[\left(\sum_{i=1}^N\cos \theta_i\right)^2\right.\right.\nonumber  \\
&& \hskip 2truecm
\left.\left.+
\left(\sum_{i=1}^N\sin \theta_i\right)^2\right]  +\beta K \sum_{i=1}^{N}\cos \left(\theta_{i+1}-\theta_i \right) \right).
\end{eqnarray}
Since two quadratic terms are present in the Hamiltonian, one has to introduce two Hubbard-Stratonovich
transformations
\begin{equation}
\exp({\displaystyle ax^2})=\frac{1}{\sqrt{4\pi a}}\int \mbox{d}z\ \exp\left({\displaystyle -\frac{z^2}{4a}+zx}\right),
\end{equation}
with auxiliary fields $\tilde{z}_1$ and $\tilde{z}_2$, obtaining
\begin{eqnarray}
Z &=& \left(\frac{2\pi}{\beta}\right)^{\frac{N}{2}}\exp\left(-\frac{\beta N}{2}\right)\frac{ N}{2\pi\beta} \int \mbox{d}\tilde{z}_1 
\mbox{d}\tilde{z}_2
\prod_{i=1}^N \mbox{d}\theta_i \nonumber \\
&& \hskip 1.5truecm\exp\left(-\frac{N}{2\beta}\tilde{z}_1^2 -\frac{N}{2\beta}\tilde{z}_2^2  +\tilde{z}_1 \sum_{i=1}^N\cos\theta_i 
+\tilde{z}_2 \sum_{i=1}^N\sin\theta_i \right.\nonumber  \\
&& \hskip 5truecm
\left.+\beta K \sum_{i=1}^{N}\cos\left(\theta_{i+1}-\theta_i\right) \right)  \\ 
&=& \left(\frac{2\pi}{\beta}\right)^{\frac{N}{2}}\exp\left(-\frac{\beta N}{2}\right)\frac{\beta N}{2\pi} \int \mbox{d}z_1 \mbox{d}z_2 
\prod_{i=1}^N \mbox{d}\theta_i\nonumber  \\
&& \hskip 1truecm \exp\left(-\frac{N\beta}{2}z_1^2 -\frac{N\beta}{2}z_2^2   +\beta z_1 \sum_{i=1}^N\cos\theta_i\right.
\left.+\beta z_2 \sum_{i=1}^N\sin\theta_i\right.\nonumber  \\
&& \hskip 5truecm
\left. +\beta K \sum_{i=1}^{N}\cos\left(\theta_{i+1}-\theta_i\right) \right)~,
\end{eqnarray}
where $\tilde{z}_i=\beta z_i$. It can be easily shown that, due to invariance under rotation in $\theta_i$,
only the integral in the modulus $z=\sqrt{z_1^2+z_2^2}$ appears leading to
\begin{eqnarray}
Z &=& \left(\frac{2\pi}{\beta}\right)^{\frac{N}{2}}\exp\left(-\frac{\beta N}{2}\right)\frac{\beta N}{2\pi} \int 2 \pi z \mbox{d}z \prod_{i=1}^N 
\mbox{d}\theta_i \exp\left(-\frac{N\beta}{2}z^2  \right.\nonumber\\
&&\hskip 2.5truecm\left. +\beta z \sum_{i=1}^N\cos\theta_i +\beta K \sum_{i=1}^{N}\cos\left(\theta_{i+1}
-\theta_i\right) \right)~.  
\end{eqnarray}
The integral over the angles has the following form
\begin{equation}
\label{eq:integrale}
\int \prod_{i=1}^N \mbox{d}\theta_i \exp\left( \sum^N_{i=1} g(\theta_i,\theta_{i+1})\right)~,
\end{equation}
where
\begin{equation}
g(\theta,\theta') = \frac{1}{2}\beta z (\cos\theta + \cos\theta') + \beta K \cos (\theta - \theta')~.
\end{equation}
Using the integral operator representation
\begin{equation}
\label{eq:transfer}
(T\psi)(\theta)= \int \mbox{d}\theta'\exp[g(\theta,\theta')]\psi(\theta')
\end{equation}
and taking advantage of the periodic boundary conditions $\theta_1=\theta_{N+1}$, integral (\ref{eq:integrale}) reduces
to \cite{solitons} 
\begin{equation}
\sum_{j=1}^{\infty}\lambda_j^N
\end{equation}
where $\lambda_j$ is the discrete set of eigenvalues of the transfer operator (\ref{eq:transfer}).
When $N$ is large, the maximal eigenvalue~$\lambda$ dominates in the sum (corrections being exponentially
small in $N$). Since the transfer operator is symmetric, all eigenvalues are real. Moreover, we expect
on a physical basis that the maximal eigenvalue be positive, because a negative eigenvalue would generate
an imaginary part of the free energy (see below) and would produce an oscillation in $N$ spoiling the
convergence to the thermodynamic limit.
Expressed in terms of the maximal eigenvalue  $\lambda$ of the transfer operator, the partition function
reads
\begin{equation}
\label{eq:partition}
Z =\left(\frac{2\pi}{\beta}\right)^{\frac{N}{2}}\!\! \!\!\exp\left(-\frac{\beta N}{2}\right)\beta N\!\!  \int\!\! z \mbox{d}z 
\exp\left(-\frac{N\beta}{2}z^2   +N\ln \lambda(\beta z, \beta K) \right)~.  
\end{equation}
Now, one can compute the partition function using the saddle-point method to perform the integral in $z$ in the
large $N$ limit. The $z$ factor in front of the integral gives a contribution which is negligible in
this limit. One gets the free energy $f(\beta)$ as
\begin{eqnarray}
\label{eq:energielibre}
-\beta f(\beta)&=& - \varphi(\beta)= \lim_{N\rightarrow\infty}\frac{1}{N}\ln Z  \\
&=&\sup_z\left[-\beta \frac{(1+z^2)}{2}+\ln\lambda(\beta z, K\beta) + \frac{1}{2} 
\ln \frac{2\pi}{\beta}\right]\\&=&
\sup_z \left[ -\tilde{\varphi}(\beta,z) \right]
\end{eqnarray}
where the rescaled free energy $\varphi=\beta f$ has been introduced for convenience, because it appears naturally
in the applicatiom of the min-max procedure \cite{Leyvraz}. In the second line, we also define the rescaled
free energy $\tilde{\varphi}$ as a function of both the inverse temperature and the Hubbard-Stratonovich auxiliary 
variable $z$.

One obtains the microcanonical entropy using the min-max method \cite{Leyvraz}
\begin{eqnarray}
s(u) &=& \sup_z \inf_\beta [\beta u + \tilde{\varphi}(\beta,z)] \\ \label{eq:ne23}
&=&\sup_z \inf_\beta \left[\beta u -\beta \frac{(1+z^2)}{2}+\ln\lambda(\beta z, K\beta) 
+ \frac{1}{2} \ln \frac{2\pi}{\beta}\right] \label{eq:minmax} \\
&=& \sup_z \left[ \tilde{s}(u,z) \right]~, 
\end{eqnarray}
where $\tilde{s}$ is defined in analogy with $\tilde{\varphi}$.
Solving the variational problem in formula (\ref{eq:ne23}) also gives the value of the microcanonical temperature 
as a function of energy $u$.

The second order phase transition line in both the canonical and the microcanonical
ensemble can be derived analytically using a perturbation theory for the transfer operator
inspired by quantum mechanics. This is possible because the magnetization $m$ varies 
continuously at the phase transition, where it vanishes. Hence all quantities can
be developed in Taylor series around $z=0$, e.g. the free energy, which is even in $z$, reads
\begin{equation}
\label{eq:sviluppp}
-\tilde{\varphi}(\beta,z) = a +b z^2+c z^4 +o(z^6)~.
\end{equation}
The change of sign of the $z^2$ term $b$ (with $c$ remaining positive) 
will indicate where the second order phase transition is located in the $(K,T)$ phase plane. The canonical
tricritical point is obtained by requiring that also $c$ vanishes.
Therefore, in order to determine the second order phase transition line in the $(K,T)$ plane, it is
enough to develop the maximal eigenvalue of the transfer operator, appearing in (\ref{eq:energielibre}),
to second order in $z$.
The series development of the operator is
\begin{equation}
(T\psi)(\theta)= \sum_i (T_i \psi)(\theta) z^i~,
\end{equation}
where
\begin{equation}
(T_i \psi)(\theta)=\int \mbox{d} \theta' \frac{1}{i!} \left[ \frac{\beta}{2} ( \cos \theta + \cos \theta') \right]^i
\exp \left(\beta K \cos (\theta -\theta') \right)\psi(\theta')~.
\end{equation}
Since we have to perform the calculation up to second order in $z$, we need to consider the perturbation
theory for the transfer operator up to second order. 
We will concentrate on the perturbative corrections to the maximal eigenvalue, considering as the
unperturbed problem the one with respect to the operator $T_0$, which can be solved exactly (see below).
Let $|r \rangle$ be the eigenvector of $T_0$ with eigenvalue $\lambda_r^0$
\begin{equation}
\label{eq:eigenvalue}
T_0 |r \rangle = \lambda_r^0 |r \rangle~,
\end{equation}
the maximal eigenvalue and eigenvector corresponding to the label $r=0$.
Non degenerate perturbation theory for Hermitian operators tells us that 
\begin{equation}
\lambda_0=\lambda_0^0+ z \langle 0|T_1| 0 \rangle + z^2 \left( \langle 0|T_2| 0 \rangle
+\sum_{r \neq 0} \frac{|\langle r| T_1 |0\rangle|^2}{\lambda_0^0-\lambda_r^0} \right) + o(z^4)~.
\end{equation}
It is well known that, introducing  $I_r$ is the modified Bessel functions of order $r$,  the eigenvalues
of $T_0$ are $2\pi I_r(\beta K)$ with eigenvector 
$\exp (i r \theta)/\sqrt{2 \pi}$ (plane waves). It is then straightforward to compute the different
terms of the perturbation series
\begin{eqnarray}
\langle 0|T_1| 0 \rangle &=& 0 \\
\langle r|T_1| 0 \rangle &=& \frac{\beta \pi}{2}(\delta_{r,1}+\delta_{r,-1})(I_0(\beta K)+I_1(\beta K)) \\
\langle 0|T_2| 0 \rangle &=&\frac{\beta^2\pi}{4}(I_0(\beta K)+I_1(\beta K))~.
\end{eqnarray}
Finally, one obtains the following formula for the maximal eigenvalue up to
order $z^2$
\begin{eqnarray}
\lambda(\beta z, \beta K)&=&2\pi I_0(\beta K)+\frac{(\beta z)^2\pi I_0(\beta K)}{2}
\left(\frac{I_0(\beta K)+I_1(\beta K)}{I_0(\beta K)-I_1(\beta K)}\right)+ ...\\
&=&A+Bz^2+....,
\end{eqnarray}
from which we can compute the coefficients $a$ and $b$ of the free energy expansion in $z$
\begin{eqnarray}
a&=& -\frac{\beta}{2}+\frac{1}{2} \ln \frac{2 \pi}{\beta} + \ln A \\
b&=& - \frac{\beta}{2}+ \frac{B}{A}~.
\end{eqnarray}
By requiring $b=0$, one gets the second order phase transition line
\begin{equation}
\frac{I_1(\beta K)}{I_0({\beta K})}= \frac{2 - \beta}{2+ \beta }~.
\end{equation}
Indeed, it is an implicit equation linking $\beta$ and $K$.
To compute the following order in the free energy, $z^4$, and the corresponding
coefficient $c$, it would be necessary to perform calculations to second order in
perturbation theory for the transfer operator. This could in principle be easily
done, but the calculations are lengthy. In order to locate the canonical tricritical 
point, we have preferred to rely upon a numerical method. Since the coefficients $a$ and $b$ 
are determined analytically, one can numerically check the difference $\tilde{\varphi}(\beta,z)-a - bz^2$ 
and observe when, for fixed $K$ and varying $T$, this difference turns from positive to negative.
In this way we get $K_{CTP}\simeq -0.171, T_{CTP} \simeq 0.267$.

The first order phase transition line is determined numerically by finding, for each value of $K$, 
the temperature $T_t=1/\beta_t$ at which the value of the free energy at the two minima in $z=0$ 
and $z \neq 0$ coincide, i.e. $\tilde{\varphi}(\beta_t,0)= \tilde{\varphi}(\beta_t,\bar{z})$, where 
$\bar{z}$ is the jump in magnetization at the phase transition.

One point of this line, the one at $T=0$, can be computed analytically. Indeed, at $T=0$ the 
free energy coincides with the energy and, therefore, one can locate the transition value of $K$
by requiring that the energy in the two phases coincides. At $T=0$ and negative $K$, one has
a transition from a ferromagnetic phase, where the mean-field term of the Hamiltonian dominates
over the short-range antiferromagnetic one, and an antiferromagnetic phase governed by the
short-range contribution. One has just to choose typical microstates corresponding to the
two phases and compute their energy. The kinetic energy can be taken at zero, since we are
at zero temperature. In the ferromagnetic phase, the magnetization vector is invariant 
under rotation; we can therefore assume that all the angles are at $\theta_i = 0$, giving
the energy
\begin{equation}
U_>=0-KN-\frac{1}{2N}(N)^2-\frac{1}{2N}(0)^2=-N\left(K+\frac{1}{2}\right),
\end{equation}
disregarding constant terms.
On the contrary, in the antiferromagnetic phase the angles are alternatively disposed at $0$ and $\pi$ 
along the lattice, leading to the energy
\begin{equation}
U_<=0+KN-\frac{1}{2N}(0)^2-\frac{1}{2N}(0)^2=+NK~.
\end{equation}
Imposing that $U_>=U_<$, one gets the transition value of $K$: $ K_{trans}=-1/4$.

The phase diagrams in both the canonical and microcanonical ensembles can be directly
derived from the properties of the functions $-\tilde{\varphi} (\beta, z)$ and
$\tilde{s} (u, z)$. 

In the canonical ensemble, the free energy is determined by
a single maximization procedure over $z$. After fixing the value of $K$, one
looks for the value of $\beta$ at which the maximum changes from $z=0$
to $z \neq 0$: this can happen in a discontinuous way for a first order phase
transition or continuously for a second order phase transition. Indeed, when
the transition is first order the local maximum of $-\tilde{\varphi} (\beta, z)$
at $z=0$ remains a local maximum. For second order phase transitions the
local maximum at $z=0$ becomes a minimum. The tricritical point is found by determining
when one changes from the first type of behavior to the other: in practice one
observes that, coming from the first order phase transition side, a local minimum
in $z \neq 0$ merges with the local maximum in $z=0$ and converts it into a local minimum.

Numerically, the crucial point is the determination of the maximal eigenvalue 
$\lambda (\beta z, K \beta)$ of the transfer operator. The integral in Eq.~(\ref{eq:transfer})
is discretized using the Bode method \cite{Bode}, which turns out to be much more
efficient than the more standard trapezoidal discretization, allowing us the calculation of the
eigenvalue with high precision using about $100$ discretization points (to obtain the
same precision with the trapezoidal method one should have worked with about $1000$ points).
The calculation of the maximal eigenvalue of the, now discrete, transfer operator is done
using a routine of the Lapack library \cite{lapack}.

In the microcanonical ensemble, the procedure is more complex because it
requires the solution of a double extremalization, see Eq.~(\ref{eq:minmax}). Moreover, one has to choose the
appropriate range of energies $u$ around the phase transition. Hence, one first
fixes the value of $u$ and performs a minimization with respect to $\beta$, and then finds
a function of $z$. The maximum of this latter function determines the equilibrium magnetization.
The search for the transition lines of first and second order is in this respect similar to
what has been already described for the canonical ensemble. The only important variant is that
for microcanonical first order phase transitions, there is a {\it temperature jump} \cite{BMR01}
at the transition energy. The two temperatures at the jump are found from the two different
values of the minimum in $\beta$ at $z=0$ and $z \neq 0$ when solving the variational problem
in $\beta$ in Eq.~(\ref{eq:minmax}). Practically, the microcanonical solution requires an order
of magnitude more of computational effort, which is due to the need of diagonalizing the
transfer integral more often.

\begin{figure}[htbp]
\centering
\includegraphics[width=10cm]{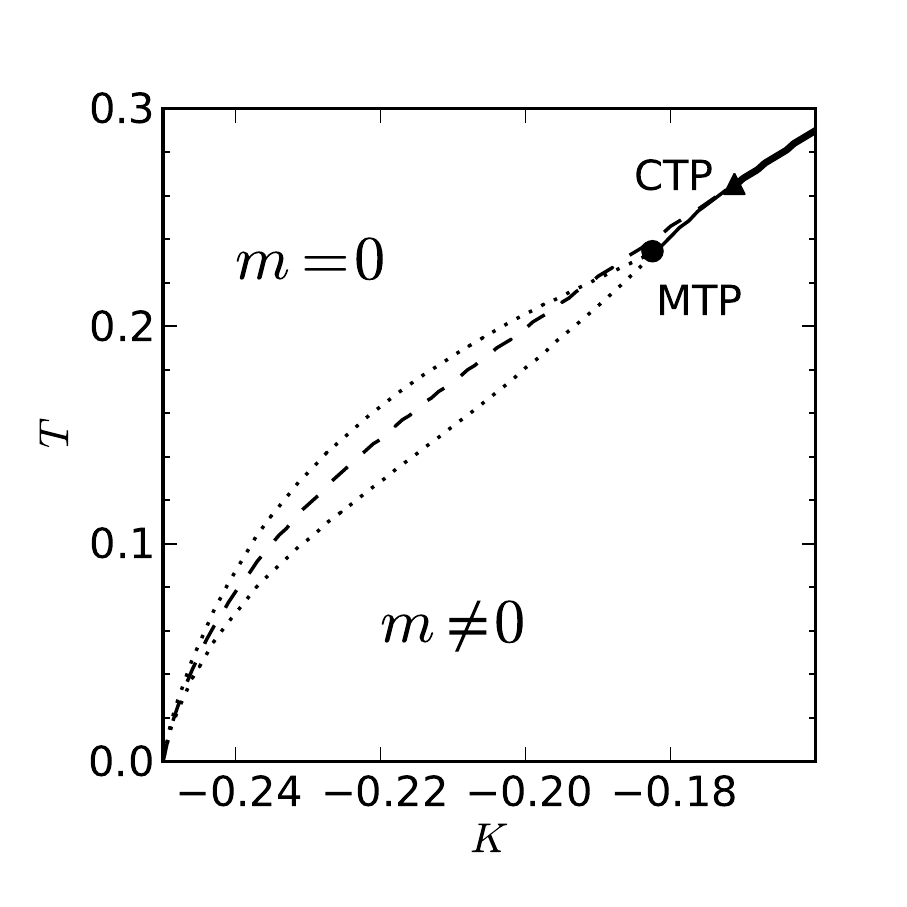}
\caption{Phase diagram of model (\ref{eq:campa}) in the $(K,T)$ plane for both the canonical and the
microcanonical ensemble. The second order phase transition in the canonical ensemble is represented 
by the thick full line. The transition changes order at the canonical tricritical point (CTP), shown by a
filled triangle, becoming first order (dashed line). In the microcanonical ensemble, the transition remains 
second order (thin full line) also beyond the canonical tricritical point, reaching the microcanonical
tricritical point (MTP). The first order phase transition in the microcanonical ensemble is represented
by the two dotted lines, as detailed in the text.}
\label{fig:dgr}
\end{figure}

The phase diagram in the $(K,T)$ plane is represented in Fig.~\ref{fig:dgr}.
For positive values of $K$ there is no competition between the short-range and
the XY mean-field term in the Hamiltonian, and therefore one expects, and finds, a second
order phase transition in both the canonical and the microcanonical ensemble (thick full line). For negative
values of $K$ there is instead competition and one can predict that for a sufficiently negative
value of $K$, the phase transition becomes first order. Indeed, we have already given a theoretical 
argument, valid at $T=0$, that suggests a first-order phase transition at $K_{trans}=-1/4$.
Hence, we can anticipate the existence of a first order phase transition line in the $(K,T)$ plane
and the presence of a tricritical point. What we find is analogous to what has been
also derived for the Kardar-Nagle \cite{nagle,kardar} Ising spin chain in one dimension
\cite{MRS05}. It is in the range of $K$ where the transition is first-order that ensemble inequivalence takes place.
As it happens also for other models~\cite{review}, the microcanonical (MTP) and canonical (CTP)
tricritical points do not coincide. In the canonical ensemble, the transition line changes to first order (dashed line) 
at the canonical tricritical point (CTP), located at $(K=-0.171, T=0.267)$ and ends at $T=0$ when $K=K_{trans}=-1/4$.
In the microcanonical ensemble, the transition is also second order on the right side of the diagram but remains 
second order also beyond the CTP (thin full line) in a region where the canonical ensemble predicts a first order
phase transition. The second order line ends at the microcanonical tricritical point
(MTP), located at $(K=-0.182, T=0.234)$. Beyond the MTP the microcanonical
transition splits into two lines (dotted lines), corresponding to the two temperatures present at the transition
energy (this is the so-called temperature jump); these lines finally merge at the zero temperature 
transition point. The two lines define a region which is not accessible in the microcanonical ensemble.
Indeed, the microcanonical first-order transition corresponds to a single line in the $(K,u)$ plane; 
it splits in the $(K,T)$ plane because different values of $T$ are found approaching the transition
energy from above or from below.

In summary, this simple spin chain system with short and long-range interactions is another example of system
for which ensembles can be inequivalent. This means that experiments realized in isolated systems, described by the microcanonical ensemble, may give
different results from similar experiments performed with well thermalized systems, for which the canonical ensemble is the
appropriate one.  It is interesting to emphasize that, although ensembles are equivalent 
for systems with short-range interactions, it is the introduction of the short-range term which has induced the inequivalence,
since inequivalence is absent in the pure mean-field XY model.

\section{A reduced model for layered spin structures}
\label{reduced}

Spin systems with both Heisenberg short-range terms and dipolar interactions have been studied by
Sievers and coworkers \cite{campionedip}. The spins are organized on $N$ layers and the interlayer
distance is sufficiently large that dipolar interaction dominates over Heisenberg exchange interaction. 
On the contrary the intralayer interaction is predominantly Heisenberg and induces a ferromagnetic coupling
between neighbouring spins within a layer. In Ref.~\cite{Layer} it has been shown that, for rod-shaped
samples, the Hamiltonian describing the system can be effectively reduced to the following one 
\begin{eqnarray}
\label{eq:effet}
H&=&\sum_{j=1}^N \frac{p_j^2}{2}-K \sum_{j=1}^{N}\cos\left(\theta_{j+1}-\theta_j\right)
+\frac{B}{2}\sum_{j=1}^N\sin^2\theta_j \nonumber\\
&&\hskip 1 truecm-\frac{1}{2N} \left(\sum_{j=1}^N\cos\theta_j\right)^2 
+\frac{1}{4N}\left(\sum_{j=1}^N\sin\theta_j\right)^2~,
\end{eqnarray}
where $\theta_j$ is an angle representing the orientation of a classical spin which describes the
macroscopic magnetization in the $j$-th layer (the modulus of such spin is constant and can be set to
unity). The constants $K$ and $B$ are determined in Ref.~\cite{Layer} from the microscopic parameters
of the sample.

Hamiltonian (\ref{eq:effet}) looks very close to the one of Eq.~(\ref{eq:campa}). However, there 
are two substantial differences: $i)$ an asymmetry in the mean field term, with the term containing 
the sines of the angles which is now positive, $ii)$ the addition of a local potential proportional 
to $B$.
Both of these terms break the rotational invariance of the Hamiltonian and make the behavior of the
system richer.
Physically, the asymmetry in the mean field term, which favours the orientation of the magnetization
vector along the $\theta_j=0$ axis (the easy-axis along the rod) is originated by the presence of
an antiferromagnetic ``demagnetizing term'' proportional to the magnetization square of the sample.
The tendency to align the spins along $\theta_j=0$ is strengthened by the local potential term when
$B>0$, and is instead opposed by the local term when $B<0$. Also the role of the nearest-neighbour
term is important, as we have seen for Hamiltonian ~(\ref{eq:campa}), since, when $K<0$, it favours 
antiferromagnetic configurations, without however fixing the global orientation of the spins.

After performing the Hubbard-Stratonovich transformation, the partition function reads 
\begin{eqnarray}
Z &=& \left(\frac{2\pi}{\beta}\right)^{\frac{N}{2}}\exp\left(-\frac{\beta N}{2}\right)\frac{\beta N}{2\pi} \int \mbox{d}z_1 \mbox{d}z_2 
\prod_{i=1}^N \mbox{d}\theta_i\nonumber  \\
&& \exp\left(-\frac{N\beta}{2}z_1^2 -N\beta z_2^2 +\beta z_1 \sum_{i=1}^N\cos\theta_i\right.
\left.+i \beta z_2 \sum_{i=1}^N\sin\theta_i \right.\nonumber  \\
&& \hskip 2 truecm\left.+\beta K \sum_{i=1}^{N}\cos\left(\theta_{i+1}-\theta_i\right) \right.
\left.-\frac{\beta B}{2}\sum_{i=1}^{N}\sin \theta_i^2 \right)~,
\label{newZ}
\end{eqnarray}
and the $g(\theta,\theta')$ function entering in the transfer matrix becomes
\begin{eqnarray}
\label{newg}
g(\theta,\theta') &=& \frac{1}{2}\beta z_1 (\cos\theta + \cos\theta') +\frac{i}{2}\beta z_2 (\sin\theta + \sin\theta')
+ \beta K \cos (\theta - \theta') \nonumber\\
&&\hskip 3truecm-\frac{\beta B}{4} (\sin^2\theta+ \sin^2\theta').
\end{eqnarray}
The free-energy can be obtained, as before, by solving the following variational problem
\begin{eqnarray}
-\beta f(\beta)&=& -\varphi(\beta) = \lim_{N\rightarrow\infty}\frac{1}{N}\ln Z \\ \label{eq:fbeta}
&=&\sup_{z_1,z_2}\left[-\beta \frac{(z_1^2+2 z_2^2)}{2}+\ln \lambda \left(\beta z_1,\beta z_2, K\beta\right) \right. 
\left. +\frac{1}{2} \ln\frac{2\pi}{\beta}\right]\\
&=& \sup_{z_1,z_2} \left[-\tilde{\varphi}(\beta,z_1,z_2) \right] ~, 
\end{eqnarray}
where $\lambda$ is the maximal eigenvalue of the transfer operator. Analogously to Eq.~(\ref{eq:ne23}), one
defines a microcanonical variational problem.

There are two additional difficulties, when determining the free energy of this model. The first one is that,
due to the lack of rotational symmetry, one has to look for extrema in two coordinates, $z_1$ and $z_2$.
The second one is that, due to the positive sign in one of the mean-field terms, the function $g(\theta,\theta')$
is complex and, consequently, the transfer operator is no more Hermitian. This latter property entails
that the maximal eigenvalue could have an imaginary part. However, this is not possible, because we have
started our calculation from a partition function which is real, and in no way it can become complex in the
end. We have then checked numerically that, indeed, the imaginary part of the maximal eigenvalue of the transfer matrix
is by twelve orders of magnitude smaller than the real part, and that its size decreases further as the number
of points in the discretization of the integral by the Bode method increases. We can thus safely assume
that the maximal eigenvalue is real. There is moreover one further simplification; by the numerical search
of the extremum for different values of the parameters, always for $T>0$ (we will come back to this point),
we have found that the non trivial maximum of $-\tilde{\varphi}(\beta,z_1,z_2)$ always appears at $z_1 \neq 0$
and $z_2=0$. This means that the symmetry is always broken is such a way that the total magnetization is 
directed along the $\theta=0$ axis. This is of course perfectly understandable when $B \geq 0$. However,
when $B<0$ the local term will tend to orient the magnetization at $\theta= \pi/2$ . What we find empirically
then means that the mean-field term always dominates at $T>0$ and determines an orientation of the magnetization
along the $\theta=0$ axis. One can thus restrict the search of the extrema of $-\tilde{\varphi}(\beta,z_1,z_2)$
to the region where $z_2=0$, which also implies that the imaginary term in (\ref{newg}) vanishes, simplifying
also the calculation of the maximal eigenvalue of the transfer operator, which is now necessarily real.

\begin{figure}[htbp]
\centering
\includegraphics[width=10cm,height=8cm]{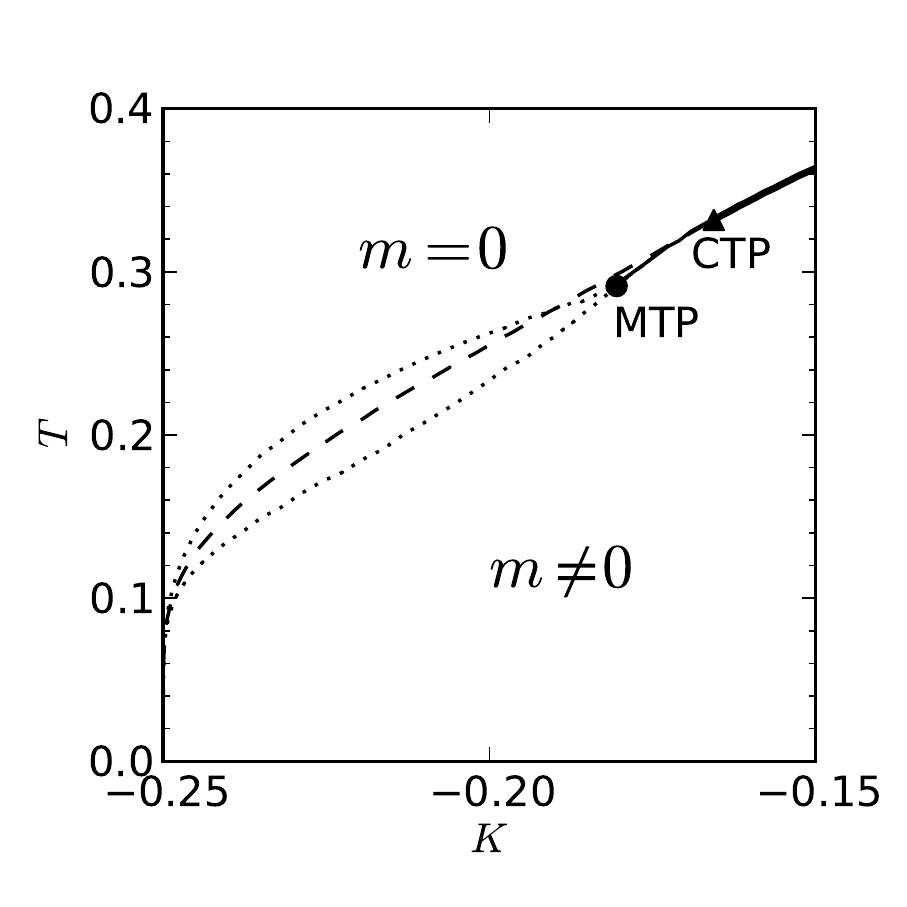}
\caption{Phase diagram of Hamiltonian (\ref{eq:effet}) for $B=0.5$. The symbols and line types are the
same as those of Fig.~\ref{fig:dgr}. The location of the canonical (CTP) and microcanonical tricritical point (MTP)
are at $(K=-0.1805,T=0.2913)$ (filled triangle), $(K=-0.1656,T=0.3319)$ (filled circle), respectively.} 
\label{bpositive}
\end{figure}

The phase diagram for $B>0$ does not show any new features with respect to the one of Hamiltonian (\ref{eq:campa}).
This can be intuitively explained by the fact that the positivity of both the local term and the mean-field
term with the sines of the angles in Hamiltonian (\ref{eq:effet}) requires, by energy minimization, that magnetization
aligns along the $\theta=0$ axis. This exactly reduces Hamiltonian (\ref{eq:effet}) to (\ref{eq:campa}). This does
not mean that one gets the transition temperatures which will be determined also by the effect
of the entropy. Indeed, at fixed $K$, transition temperatures increase as $B$ is increased, and the tricritical
points move towards higher values of $K$. However, qualitatively, we do not expect significant changes of the 
phase diagram. Moreover, the transition at $T=0$ happens again at $K_{trans}=-1/4$ for all values of $B>0$, 
as can be easily proven by repeating the argument given in Section~\ref{HMF}. In Fig.~\ref{bpositive}, we display 
the phase diagram of Hamiltonian (\ref{eq:effet}) for $B=0.5$ in both the canonical and the microcanonical ensemble. 
The transition lines and the tricritical points are drawn with the same line types and symbols as those of 
Fig.~\ref{fig:dgr}. In this positive $B$ domain, the phase diagram is essentially similar to what was obtained for model~(\ref{eq:campa}).

\begin{figure}[htbp]
\centering
\includegraphics[width=10cm,height=8cm]{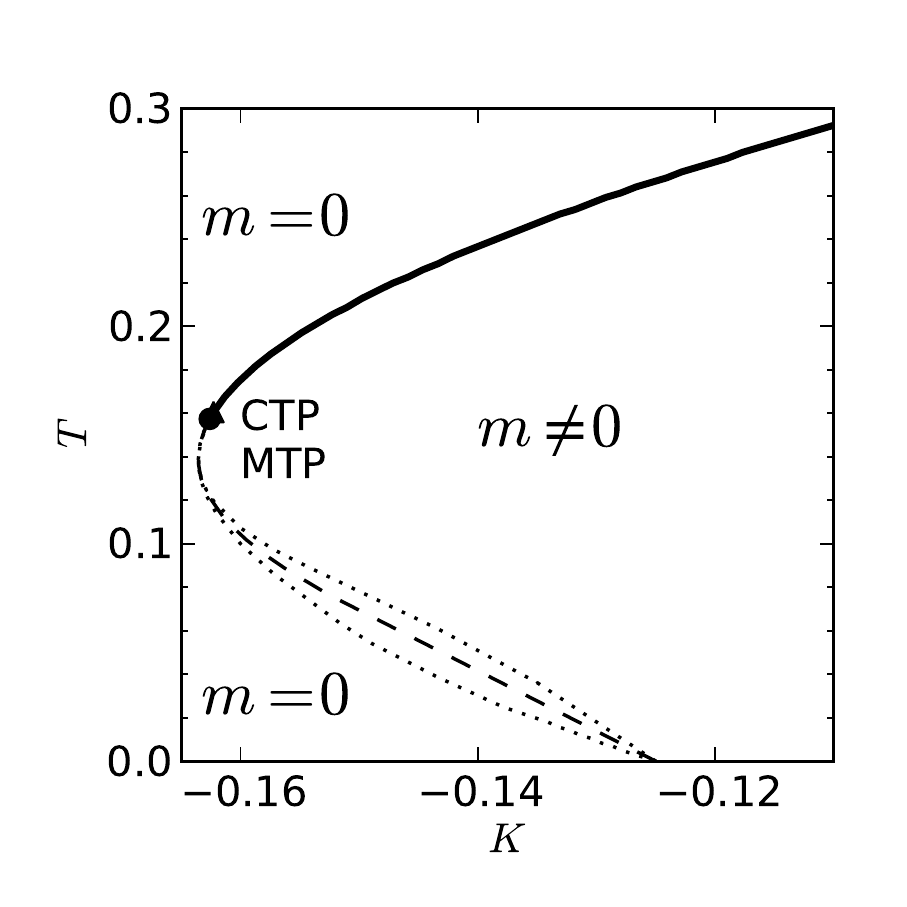}
\caption{Phase diagram of Hamiltonian (\ref{eq:effet}) for $B=-0.5$. The symbols and line types are the
same as those of Fig.~\ref{fig:dgr}. The location of the canonical (CTP) and microcanonical tricritical point (MTP)
are at $(K=-0.1623,T=0.1605)$ (filled triangle), $(K=-0.1626,T=0.1574)$ (filled circle), respectively.}
\label{bnegative}
\end{figure}

The case $B<0$ is instead more interesting. One observes numerically that, not only the transition
temperatures, as expected for reason of symmetry with respect to the previous $B>0$ case, reduce (at fixed $K$) as
$B$ becomes more negative, but that the transition lines bend, with the lower part of the transition
lines themselves which slides towards larger values of $K$. 


\begin{figure}[htbp]
\centering
\includegraphics[width=10cm,height=8cm]{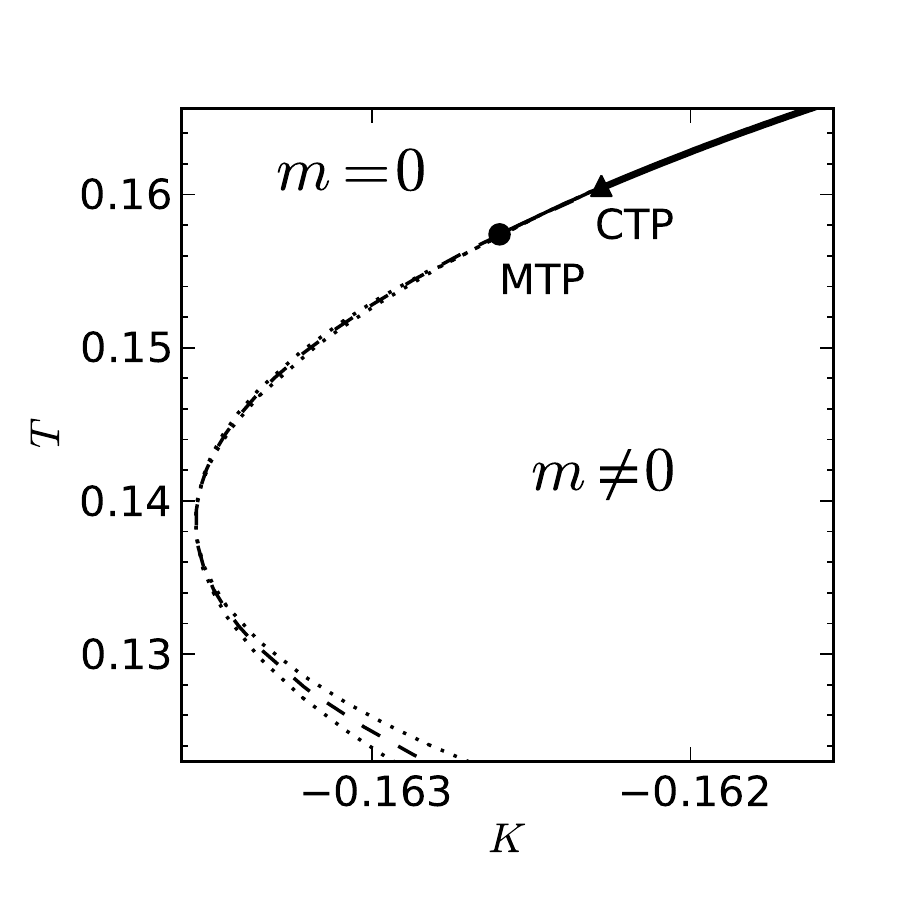}
\caption{Zoom on figure \ref{bnegative} around the canonical and microcanonical tricritical points.}
\label{bnegative-zoom}
\end{figure}

The transition at $T=0$ is also shifted towards a larger value of $K$. Can this be
understood? The only way this can be justified is that the local term, proportional to $B$, plays a role. This can
happen, at $T=0$, only if the antiferromagnetic zero-magnetization state has its spins oriented along
the $\theta=\pi/2$ axis. Let us then repeat the zero-temperature argument of the previous section
for the sake of clarity. In the magnetized phase, the mean-field terms dominate and the spins will be all 
oriented at $\theta_i=0$ or $\pi$. The corresponding energy will be
\begin{equation}
\label{eq:label1}
U_>=0-KN-\frac{1}{2N}(N)^2+\frac{1}{4N}(0)^2+\frac{B}{2}(0)=-N\left(K+\frac{1}{2}\right).
\end{equation}
In the zero-magnetization phase instead, the local and nearest-neighbour terms in the Hamiltonian
will dominate, and the microstates will be antiferromagnetic, with the spins alternatively oriented at
$\pi/2$ and $-\pi/2$. This gives an energy
\begin{equation}
\label{eq:label2}
U_<=0+KN-\frac{1}{2}(0)^2+\frac{1}{4N}(0)^2+\frac{B}{2}(N).
\end{equation}
Equating the two energies, $U_>=U_<$, one obtains $K_{trans}=-(B+1)/4$. This value is in perfect agreement
with the one determined numerically. 

For $B$ sufficiently negative, e.g. $B=-0.5$ as in Fig.~\ref{bnegative}, one observes the curious phenomenon
of {\it phase reentrance}:  a vertical line
in the $(K,T)$ plane crosses twice the transition line.  By lowering the temperature at a fixed value of $K$, one 
observes  first a disorder/order transition and then, by further lowering the temperature an order/disorder transition.  
Fig.~\ref{bnegative-zoom} presents a zoom of this interesting region. Interestingly, the microcanonical 
first-order phase transition lines (dotted) seem to coalesce at the bending point before separating again below it. 
The system is disordered down to zero temperature. However, in our case, as we have commented above, the
microstates close to zero temperature, although they have a zero magnetization, are antiferromagnetic
states.

Phase reentrance was first described by Griffiths and Wheeler \cite{fasirientranti} and has found later
several experimental verifications for colloids and polymers. Recently, it has been also discovered
in the context of out-of-equilibrium phase transitions of the HMF model \cite{staniscia}. 

A different interpretation of the phenomenon of phase transition line bending that we observe can be
given in terms of the classification of phase transitions for long-range systems described in Ref.~\cite{barrebouchet}.
Let us refer for example to the situation represented in Fig.~\ref{bnegative-zoom}. When increasing $K$
from negative values one passes from a situation where no phase transition is present (one is always
in the $m=0$ phase for all temperatures) to one where two first-order phase transitions are present in
both the canonical and microcanonical ensemble. This is a case of {\it azeotropy} with ensemble inequivalence
discussed in Ref.~\cite{barrebouchet}. To our knowledge, this kind of azeotropy has never been found. 
A different case of azeotropy at a second-order phase transition, i.e. with ensemble
equivalence, was indeed found for certain two-dimensional flows in Ref.~\cite{venaillebouchet}. This type of
phase transition pattern could be present in our model for smaller values of $B$, when presumably both
the canonical and microcanonical tricritical point lie below the bending point of the transition line.
Moreover, a case of second-order azeotropy with ensemble inequivalence could be found when the canonical
tricritical point is above the bending point and the microcanonical one is below. All this remains to
be ascertained. 

Among the curious effects related to phase reentrance is the fact that, even in mean-field models, one can have
non classical exponents at second order phase transitions. This would happen in our model in the latter case
we have discussed in connection with second-order azeotropy, in which the value of $K$ is tuned in such a way 
that a vertical line in the $(K,T)$ plane touches tangentially the transition line of second order.

\begin{figure}[htbp]
\centering
\includegraphics[width=10cm]{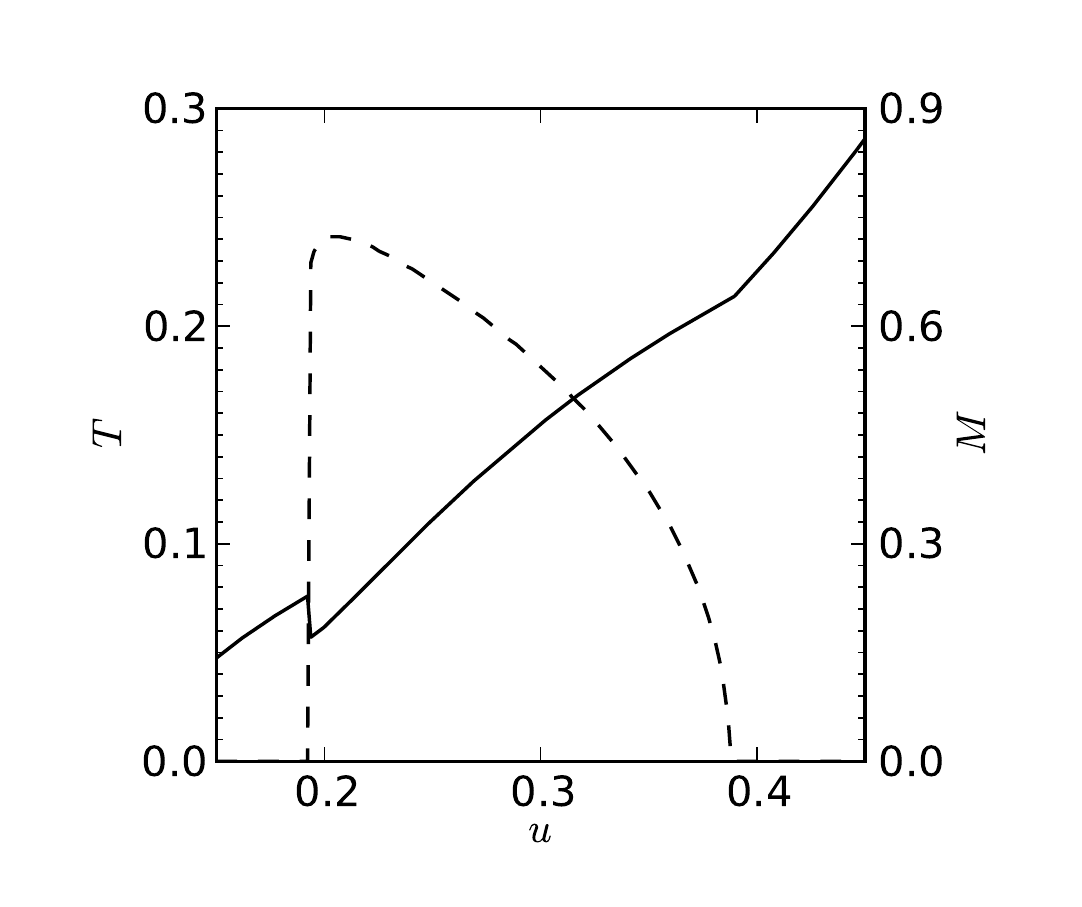}

\caption{Caloric curve and magnetization vs. energy at $B=0.5$ and $K=-0.15$. Microcanonical temperature 
(full line) is plotted vs. energy (left axis) showing first a temperature jump, corresponding to a first order microcanonical 
phase transition, followed by second order phase transition detected by a change in slope. Magnetization values
(dashed line) are shown on the right axis.}
\label{caloric}
\end{figure}

The phase reentrance phenomenon is, in our model, more and more pronounced as $B$ becomes more negative.
However, if $B$ is so negative that the phase transition point at $T=0$ moves to positive $K$, we expect
a change in the nature of the phase transition, because the microstate cannot be antiferromagnetic in this
case. By varying the value of $B$, different patterns of phase transitions can be observed. Increasing temperature
one can have a first order phase transition followed by a second order one, as shown in Fig.~\ref{caloric}.
Otherwise, one could first meet a first order phase transition and then a second order one or, finally, meet
two second order phase transitions.

\section{Conclusions}
\label{conclusions}

In this paper we have studied in full details the phase diagram of a Hamiltonian
with local, nearest-neighbour and mean-field couplings which is obtained from a
microscopic model of dipolar layered magnets, e.g. 
$(C H_{3}NH_3)_2CuCl_4$ \cite{campionedip,Layer}.
 
Such layered magnets share the feature that the spins (here the $Cu$ atoms) are 
organized on planes, and below a given temperature they order ferromagnetically
on the plane.
However, since the distance between layers is such that the Heisenberg exchange
interaction becomes negligible with respect to dipolar interactions, the global
ferromagnetic order is reached at much lower temperatures. For a sample of
ellipsoidal form, the dipolar interaction gives rise to a term in the energy which
is proportional to the square of the total magnetization, with a coupling constant 
which depends on the geometric features of the sample. This heuristically explains
why in our effective Hamiltonian (\ref{eq:effet}), we have both a term describing
the interaction between nearest-neighbour ``classical" XY spins, which represents the
magnetization in a layer, and mean-field terms which take the dipolar interaction
between all the spins in the sample into account.

Ising models with both short-range and long-range terms in the Hamiltonian have been 
independently considered \cite{nagle,kardar,MRS05} and studied in both the canonical
and microcanonical ensemble. They show all the typical features of systems with
long-range interactions \cite{review,longrange,assisi,leshouches}, i.e. ensemble 
inequivalence, negative specific heat, temperature jumps, breaking of ergodicity 
and quasi-stationary states. Predictions in the microcanonical ensemble could be
checked in experiments performed in conditions of thermal isolation or fast 
variation of external parameters (temperature quenches, external magnetic
field sweeps) \cite{Barbara}.

Besides the Ising models mentioned above, XY models on one-dimensional lattices
with both short and mean-field terms have been studied \cite{Campa}. A preliminar
reconstruction of the phase diagram and a study of the dynamics have been performed.

With the aim of looking for experimental realizations of long-range effects, in 
Ref.~\cite{Layer} an attempt has been made to fill the gap between the Hamiltonians
introduced for magnetic layers dipolar systems and the XY Hamiltonians studied
in Ref.~\cite{Campa}. This  has led us to study Hamiltonians of the type of
Eq.~(\ref{eq:effet}).

A new feature of this Hamiltonian, with respect to the previously studied 
XY Hamiltonian, is the presence of anisotropies. The first one is due to a
different coupling (even in sign) of mean-field terms in cosines and sines
of the angles of the XY rotor. The second is due to a local term, proportional
to the coupling $B$, analogous to an external magnetic field. The value of
this coupling constant $B$ is calculable from the geometric and physico-chemical
properties of the sample and turns out to be positive. However, in order to
perform the analysis in full generality, we have considered both positive and
negative values of $B$.
 
The solution of Hamiltonian~(\ref{eq:effet}) is possible via a smart combination
of the {\it min-max} method introduced in Ref.~\cite{Leyvraz} with the transfer 
operator method \cite{krum,Aubry,solitons}. The diagonalization of the transfer operator
has been performed using the Bode discretization \cite{Bode}, which allows to
work with substantially smaller matrices with respect to the more standard trapezoidal
discretization. These methods allowed us to obtain a greater precision in the determination
of the phase diagram with respect to previous studies of Hamiltonians of this kind.

We have found that for $B>0$ the phase diagram is qualitatively similar to the one
of the XY Hamiltonian (\ref{eq:campa}). In the canonical ensemble a second order
phase transition is found when the nearest-neighbour coupling $K$ is positive. The 
phase transition becomes first order if $K$ is sufficiently negative, inducing 
antiferromagnetic patches in the microscopic spin configurations (although antiferromagnetic
global order is not possible in one dimension). The second order phase transition line 
is separated from the first order one by a tricritical point. In the microcanonical
ensemble, the second order phase transition line coincides with the one of the 
canonical ensemble until the canonical tricritical point is reached. Beyond this
point, the phase transition remains second order in the microcanonical ensemble
until the line reaches a microcanonical tricritical point, which differs from the
canonical one. At the microcanonical tricritical point the transition line splits in two,
because of the presence of a first-order microcanonical transition. The two
lines join again at zero temperature. These features are common with those found for
the phase diagram of other systems with long-range interactions and are an impressive
manifestation of {\it ensemble inequivalence}.

Definitely more interesting is the $B<0$ case. The phase transition lines bend at
lower temperatures as $B$ becomes more negative until the lines take a {\it boomerang}
shape. This is a signature of the phenomenon of phase reentrance, first discussed
by Griffiths and Wheeler \cite{fasirientranti}. When lowering the temperature at
fixed $K$ the system first go through a disorder/order transition and then again
through an order/disorder one and is, counter intuitively, disordered down to
zero temperature. An alternative interpretation of this phenomenon makes reference
to {\it azeotropy} \cite{barrebouchet,venaillebouchet}. The two transitions can be of first or second order depending
on the chosen value of $B$. In the microcanonical ensemble, also the splitted
first-order phase transition lines bend. This latter effect determines the
presence of a different {\it caloric curve} which shows a temperature jump
followed by a more standard change in slope at the second order phase transition.

Much remains to be done in the study of this kind of models. One could mention
that the solution of a more complicated Hamiltonian, which takes into account also
the off-plane motion of the spins \cite{Layer}, is possible using the same methods.
The additional difficulty is that one has to diagonalize a transfer operator
containing two angles and this makes the application of standard numerical methods
quite hard.

Fully absent in this work is the study of dynamical effects. Both breaking of
ergodicity and quasi-stationary states should be present in Hamiltonian (\ref{eq:effet})
and would be certainly interesting to investigate.

\ack
We thank A. Campa and B. Barbara for useful suggestions and discussions. SR thanks UJF-Grenoble and ENS-Lyon
for financial support and hospitality. SR acknolwledges the financial support of the COFIN07-PRIN
program ``Statistical physics of strongly correlated systems at and out of equilibrium'' of the Italian MIUR
and of INFN. PdB would like to thank P.~Gaspard for support. The research 
of PdB is financially supported by the Belgian Federal Government (Interuniversity Attraction 
Pole ``Nonlinear systems, stochastic processes, and statistical mechanics'', 2007-2011).

\end{document}